# From Personal Data to Digital Legacy: Exploring Conflicts in the Sharing, Security and Privacy of Post-mortem Data

Jack Holt
Open Lab, Newcastle University
Newcastle Upon Tyne
j.holt3@newcastle.ac.uk

James Nicholson
Northumbria University
Newcastle Upon Tyne
james.nicholson@northumbria.ac.uk

Jan David Smeddinck
Open Lab, Newcastle University
Newcastle Upon Tyne
jan.smeddinck@newcastle.ac.uk

## ABSTRACT

As digital technologies become more prevalent there is a growing awareness of the importance of good security and privacy practices. The tools and techniques used to achieve this are typically designed with the living user in mind, with little consideration of how they should or will perform after the user has died. We report on two workshops carried out with users of password managers to explore their views on the post-mortem sharing, security and privacy of a range of common digital assets. We discuss a post-mortem privacy paradox where users recognise value in planning for their digital legacy, yet avoid actively doing so. Importantly, our findings highlight a tension between the use of recommended security tools during life and facilitating appropriate post-mortem access to chosen assets. We offer design recommendations to facilitate and encourage digital legacy planning while promoting good security habits during life.

## 1 INTRODUCTION

As digital technologies and ubiquitous computing become more commonplace, we are increasingly defining ourselves by aspects of our digital lives. To protect our digital data and identities, there is a need for additional security mechanisms and behaviours, such as password managers, which help prevent common opportunistic attacks, and multi-factor authentication, which adds a second layer of protection to digital accounts. As such security measures become more complex, there arises a question of how the data and identities protected with these mechanisms are managed after the death of the individual. This is particularly evident in the case of social media technologies, which have an immediate and obvious need to ensure that accounts representing deceased individuals are retired. For example, it has been predicted that, with continued uptake, the quantity of deceased Facebook users would number in the billions before the end of the 21st century [37]. However, much less visible to others are the digital possessions and accounts that are not public-facing and the value of which, whether financial or sentimental, is at risk of loss upon the owner's death.

Presently, the act of making comprehensive plans for this digital legacy is hampered by a lack of technological knowledge or means and limited established social practices. Without an individual having made such plans, their loved ones are unlikely to be aware of the existence and whereabouts of all of their meaningful or valuable digital assets, and may be left unable to access those that they do know of. In cases where access is possible, they may be unaware of the deceased's preferences in terms of the continued privacy of their data, and face the uncertainty of deciding when it is and is not appropriate for them to access password-protected, or otherwise private, information.

This study set out to begin understanding how technology might be used to encourage and enable security-aware technology users with active digital lives to make plans for their data and online accounts in the event of death. Through two workshops, users of password managers generated digital legacy plans and discussed privacy implications for a variety of individual online accounts and data. Using thematic analysis, this paper explores how people think about the privacy and value of their data as it exists during life and after death, and how the security of, and access to, such data might change after a person has died. We highlight a potential for increased difficulty in carrying out such plans when an individual exhibits strong security behaviours. Building on the proposed plans and obstacles identified by the participants, we present recommendations for the design of digital legacy functionality such as that found in password managers to more widely encourage this nature of planning and to ensure that legacy plans remain unimpeded by strong security habits. In doing so, this research seeks to contribute towards an understanding of individual reasoning regarding digital legacy planning, and represents an early step towards improved control over personal privacy, cybersecurity and data management.

## 2 BACKGROUND AND RELATED WORK

### 2.1 Finding Meaning in Data

The extent to which digital items might be considered to constitute part of one's personhood or self is a subject with a rich research history, a prominent example of which is the work of Russell Belk. In 1988, prior to the advent of the internet, Belk argued that one's possessions can form an integral part of how a person constructs their sense of identity and formulates the narrative of their life - as he phrased it, their "extended self" [4]. Twenty-five years later, he reconsidered this in the context of the notable changes brought about by the internet and ubiquitous computing, generating a number of key principles of how such things might affect the "extended digital self" [5]. In doing so, Belk identified five categories of departure from the original Extended Self argument that have been brought about by digital advancements and resultant social changes. Digital entities may have no physical representation, what he referred to as *dematerialisation*. New technologies further offer *reembodiment*, with avatars or personas used to represent their identity. The nature of how and what people *share* about themselves changes, along with how an individual can control sharing of personal information by others. Social media and other communication tools offer *co-construction of self*, with important life moments or media able to collect a digital patina through interactions such as messages and comments. Finally, digital entities can represent a *distributed memory* of life histories, memories and personal narratives.

When considering traditional components of the extended self, sentimental value can become attached to simple and common objects, which in turn presents an opportunity for some sense of self to be transmitted from one person to another through the gifting of meaningful objects. Spence [42] considers how this process of giving something with attached or embodied meaning – what she terms "inalienability" – might be found or replicated in the context of digital possessions. Digital items, unlike physical items, can often be duplicated and could be shared with any number of recipients. What gives them meaning, and forms a distinction between gifting and sharing, is the thought process behind why a specific digital thing might be given to this particular recipient. Objects with inalienable traits tend to have attached meanings, histories or narratives, and the process of gifting such objects to another individual is personal in nature. In this, Spence creates a distinction between relational gifting (e.g. a one to one gift) and transactional gifting (e.g. sharing a file with a group). Bringing this line of thinking into the context of digital legacy, there is a need for viable options that support not only the distribution of valuable or important assets, but also the gifting of valued parts of the self in a way that fosters more relational and inalienable bequests.

### 2.2 Management and Curation of Data

With the proliferation of new forms of digital activities and accumulation of associated data and credentials, effective management of digital assets can become a burden. Prior research has indicated that most people take actions to store information for future use, with a tendency to overkeep data [53]. Technology users can typically be placed along a spectrum ranging from minimalistic tendencies of data retention and organisation practices towards digital hoarding, in which cases all data is kept in case it one day proves to be of importance [34, 47]. Digital hoarding, as defined by van Bennekom et al, is "the accumulation of digital files to the point of loss of perspective, which eventually results in stress and disorganisation" [46]. In the context of leaving a comprehensible digital legacy, such behaviours may be detrimental to one's ability to engage with a planning process and express sufficient control over what happens to that data after death.

### 2.3 Protecting Digital Data

The most common method for protecting data and accounts is using alphanumeric passwords. Over the past decade, it has become clear that passwords do not meet modern demands, however, better alternatives are yet to be implemented [20]. A well-known limitation of passwords is their capacity to be shared (or the inability to do so securely). For example, typical advice discourages users from sharing passwords, as this can compromise the security of their accounts [13]. However, prior work has shown that passwords are commonly shared due to need [41], convenience [38], or for social credit [11].

Another key limitation of passwords involves human cognitive limitations. While remembering a handful of passwords may be feasible, users are currently required to remember passwords for an average of 26 regular accounts [39], although the number of accounts could easily reach hundreds when considering less frequently visited accounts. In order to manage this, people take shortcuts in their password usage, typically resulting in a negative impact on the security of systems. Workarounds include reusing passwords across accounts [43, 52], which risks an account compromise leading to other accounts being compromised [22].

It is becoming common for people to use mobile devices for browsing the internet. This means using passwords on these devices to log into accounts, yet touchscreen mobile devices make the creation and use of complex passwords frustrating and less likely [29], compounded by the fact that users already fail to create strong-enough passwords [45]. These limitations clearly signal the need for more usable solutions. Because of these issues, academics, professionals, and governments encourage users to take up extra security measures (e.g. [2]). These include password managers – software that can generate complex unique passwords, and, more importantly, store them so that users do not need to remember them.

### 2.4 Digital Identity Management

A consequence of using tools for the tracking and storage of online credentials is that, through the collection of the various aspects of one's accounts, a representative picture of their digital footprint – or part of it – is formed in a single location. Password managers present to the user a rare opportunity to peruse, manage and review this representation, often with features for sorting those accounts into folders, annotating them, sharing selectively with others and assigning contacts to be granted emergency access privileges. As such, they may be seen as a source of metadata regarding a person's online assets, irrespective of their capacity to provide authenticated access to those resources. Prior research has considered the potential value of metadata in understanding and engaging with large quantities of accumulated digital information, and how such data may be meaningful to others, including generations to come [17]. Others have considered how, for some individuals, such as transgender people, there may be strong personal reasons to curate or otherwise express control over digital identity [18].

Increasingly, the distinction between a person's data and their online credentials can become blurred. Recent design work has confronted the challenges behind keeping and discarding personal data in a "post-cloud" context, wherein data is not considered to be limited to a device or location, but exists in some combination of cloud and devices [48]. Where data is tied to a person's identity rather than to the physical device it is stored on, the challenge of maintaining order becomes pressing. Their study found that some participants favoured individual control over what is kept or deleted, and others preferred automated alternatives. Ultimately, they derive themes that highlight personal responsibility in managing one's own data, the burden of doing so, and the importance of context in finding appropriate mechanisms of personal data curation.

## 2.5 Data after Death

The development and discussion of effective mechanisms of defining and delivering a working digital legacy plan has been considered by a range of HCI research and commercial products and services (e.g. [21, 25, 40]). These include specially-designed estate planning services, letters of instruction attached to an individual's last will and testament, software designed specifically for digital legacy management [15] and the use of password managers as a means of directly passing credentials [8]. It has been noted that there are distinctly opposed views in respect to how this ought to be done, particularly regarding preferences towards deletion or towards preservation [26], however in either case there has been little success in designing approaches that are both technologically feasible and sufficiently accessible for the general population to engage with the process. Brubaker et al. emphasise the burden of responsibility placed on the recipients of digital inheritance, and propose that instead of considering such individuals as inheritors or beneficiaries, the term *steward* is more appropriate [7].

In addition to logistical concerns relating to the transmission of passwords and data, there are also broader changes that should be considered regarding inheritance and dying in the digital age. These include novel approaches to designing for digital remembrance [3], digital archiving of sentimental objects [23], new ways of social mourning [15, 51], the stakeholders involved immediately following death [32], and the extension and preservation of personhood through technology [28, 50]. Underlying this are questions regarding what is the nature of privacy after death, which is typically not treated with the same reverance as privacy during life [12]. This also relates to a further question of how digital remains should be treated, with some arguing for post-mortem data to be considered as "the remains of an informational human body" and that the handling of such data should be considered under ethical frameworks similar to those used in disciplines such as archaeology [37], and others going as far as to declare a moral duty for the preservation of such artefacts [44]. Some argue that informational remains are more sensitive than the human body itself, and protections that exist for the body after death (e.g in relation to organ donation) should be extended to include digital elements [9].

The concept of privacy after death is sometimes known as post-mortem privacy. Our usage of the term is in keeping with the usage of Edwards & Harbinja, who use it to describe "the right of a person to preserve and control what becomes of his or her reputation, dignity, integrity, secrets or memory after death" [12]. However, while it may be considered a right by some, data protection regulations typically apply only to living individuals [19]. Furthermore, attempted post-mortem access to data that is stored according to the Terms of Service of an online provider may prove difficult, as many prohibit user assignment of rights and do not allow users to share login credentials [31]. Even in cases where service providers make some allowance for inheritance of user data, the opportunities are limited. In the case of Facebook, the process relies on the user naming their legacy contact prior to death and allows no assignment of rights if this has not been done [30]. While legal innovations are underway in some countries (e.g. RUFADAA in the US [49], the Digital Republic in France [1]), many challenges exist for those attempting to legally access personal data associated with a deceased individual and it remains a legal grey area [30].

A recent survey of technology users in Israel reveals a range of personal opinions on whether access should be granted to personal email, social networking and cloud storage accounts after death, with 45% - 50% of those surveyed indicating a preference for someone to have full access to their accounts, 31% - 36% expressing that all access should be denied, and the remaining approximate one-fifth showing a preference for some degree of partial access [33]. An implication of this division of opinion is that without explicit post-mortem privacy preferences, most next of kin are unlikely to be in a position where they can safely assume the appropriate course of action regarding an individual's digital remains. Furthermore, security mechanisms may facilitate or prevent access to data in unintended ways.

It is also important to acknowledge that privacy and the appropriateness of access to data is not always clear cut and depends heavily on situational and relational contexts. The concept of "contextual integrity", as developed by Helen Nissenbaum, rejects the notion that some things are implicitly private and others are not [35]. Rather than some data or information being sensitive and others not, or some data being private and some public, contextual integrity considers the flow of data, and how appropriate that flow of data is given its context. For example, whether it is appropriate for data to be transferred to a certain recipient, and further, when is that transfer appropriate. Viewing post-mortem privacy from this contextual perspective illustrates the potential complexities that are involved in attempting to plan one's digital legacy, and the difficulties that may interfere with adequately defining privacy preferences. Our research takes a qualitative approach, seeking to identify some of the reasoning behind granular post-mortem privacy preferences, how they might be defined and executed, and what are some of the main barriers to comprehensively specifying such intentions.

## 3 METHODOLOGY
### 3.1 Overview
Our study is based on discussions from two workshops, in which participants were asked to collectively decide on a post-mortem plan for a variety of common forms of data and online accounts, before discussing the challenges of digital legacy more broadly.

### 3.2 Participants

The use of password managers was considered to be essential criteria for participation. In restricting participants to those who use password managers, this recruitment process was intended to increase the likelihood of attendance by individuals with active digital lives and who are security and privacy conscious. In order to locate appropriate participants, the research was advertised within the School of Computing at Newcastle University. In inviting students of the School of Computing, there was anticipated to be a higher incidence of computer and technology literacy and interest. It is important to acknowledge that this is by no means a representative sample of the population, but rather a targeted and deliberate selection of individuals who are likely to be engaged and informed about relevant online technologies. Participants were invited to attend if they used any type of password manager, and offered a small shopping voucher as a token of gratitude for their attendance.

A total of 14 participants took part in the research. Two hour-long workshops were held, each attended by 7 participants. The first workshop consisted of 4 males and 3 females, and the second consisted of entirely males. All participants were connected with computer science, either at undergraduate, postgraduate or post-doctoral level.

### 3.3 Pre-attendance questionnaire

Prior to attending the workshops, participants were invited to take part in a short, optional questionnaire. No identifiable data was collected during this questionnaire, and it was treated as a priming activity and a way for the researcher to have an awareness of the typical password manager usage patterns of those attending the workshops. The questionnaire revealed that participants used a range of password managers, and 8 were current or past users of Lastpass. The number of stored credentials per person ranged from 25 to 650, with half of the respondents reporting at least 100. 4 respondents reported taking regular action to review the contents of their password manager, including sorting into folders, deactivating old accounts and updating passwords. A variety of reasons were offered for their use of password managers, including improved security, convenience and as a memory aid.

### 3.4 Workshop materials

With the aim of exploring how password managers may be seen as gateways to a person's digital self, these workshops were designed to exemplify a wide spectrum of digital entities and generate discussion around the privacy and meaning associated with each. Digital accounts, data and files that both would and would not be likely to be represented within a password manager were included as part of this. These were printed on cards, forming a deck of 30 cards in total. Cards included well known social networking sites, such as Facebook and Twitter; common digital tools such as email and cloud storage; online services used for activities such as banking, gambling and dating; digital purchases/leases such as e-books, video games and music streaming services; individual files or folders that might be stored locally or on a cloud service; and two physical devices - the personal computer and the mobile phone.

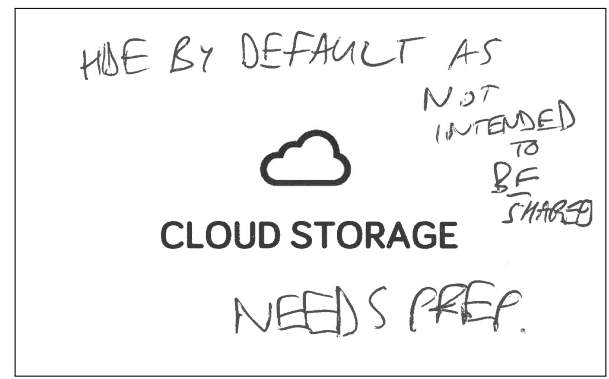

**Figure 1: An annotated digital entity card**

### 3.5 Procedure

In order to encourage all participants to contribute to the conversation, and promote orderly and balanced discussion, the cards were used as part of a turn-based activity. The process for doing so was for each participant in turn to draw a card and read it aloud, and for the group to discuss whether the card should be placed under either *plan*, *hide* or *ignore*, with the person who drew the card to make a final decision on behalf of the group. Cards filed under *plan* represented items for which there ought to be a plan in place relating to access, ownership, sharing or deletion in the event of the owner's death. Those placed under *hide* were seen as fundamentally private and their existence should be kept secret and deliberately excluded from any plans. All other cards were to be placed under *ignore*, signifying no strong feelings about the fate of those assets in the event of death.

After approximately 45 minutes of discussion prompted by this card activity, the moderator turned the conversation to digital legacy more broadly. As it was unclear beforehand how long participants would take to arrive at a decision for each card, more had been produced than were expected to be required. In both workshops, cards frequently prompted long discussions and debates, and ultimately only a portion of the provided cards were discussed (16 and 9, respectively).

### 3.6 Analysis of the workshops

The audio from the workshops was transcribed and analysed inductively using thematic analysis following Braun & Clarke [6]. The data was not analysed with an aim of seeking an objective truth or reality about the nature of privacy and the importance of data, but rather in an attempt to understand the individual views of the participants and how their values and needs might affect the challenge of secure digital legacy planning. Both workshops were coded in their entirety, and considered in terms of the meanings and values expressed by participants as part of their decision-making process as well as the specific decisions and suggestions that they arrived at. These are presented as individual sections in this paper, with the Findings section presenting the themes constructed from across the data set, and the Design Considerations section identifying some specific ideas and suggestions which are not necessarily found throughout the data but rather in specific instances, such as in discussions about how email accounts should be handled or what the role of password managers is. Presented in the **Findings** below are three themes derived from the thematic analysis: *Death as a privacy transition point*; *the post-mortem privacy paradox* and *it's a bit of a grim thing to be doing*. The subsequent **design considerations** section can best be understood as a curated selection of practical issues, barriers and solutions identified by our participants when challenged to plan for a given digital entity.

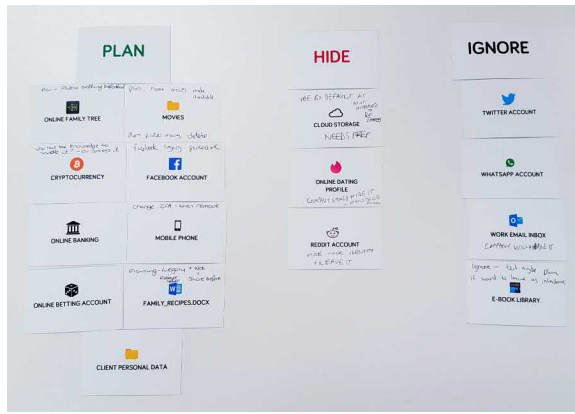

**Figure 2: The final layout of the 16 cards discussed in workshop 1.**

### 3.7 Ethics

Research that involves human participants in any way requires careful ethical consideration, and research that involves considerations of mortality and bereavement should be approached with sensitivity [27]. To do so, the research process was designed to avoid direct confrontation with personal experiences of this nature at this formative stage. The workshops were therefore designed in a way that prompted group discussion about primarily hypothetical considerations, leaving individuals with the freedom to offer personal insights and experiences only when they felt comfortable doing so.

Participants who identified as having recently experienced a bereavement, were suffering from a serious illness or who were otherwise likely to be upset or distressed by discussions relating to death were advised not to take part. A second moderator was present at both workshops in order to assist with any unexpected participant distress. Participants were advised that the topic of the research involved what happens to our digital assets when we die on multiple occasions: during advertisement of the research; via email prior to attendance; in person on their attendance at the workshops and on paper via the information sheet and consent form. They were further aware that they could withdraw from the research at any time without penalty.

The research described in the following sections was conducted following ethical approval via Newcastle University's ethics procedure.

## 4 FINDINGS

### 4.1 Death as a privacy transition point

In considering the plans that are, or are not, necessary for the wide variety of accounts and forms of data presented to them, the first consideration for any given digital entity tended to be an assessment of its inherent privacy. The logical approach to this was frequently to seek continuation of the privacy granted to that entity during life; the privacy level attributed to it can then be treated as a reflection of intended privacy after death.

> I find it an interesting thing, because, if you look at your email account right now, you would assume that nobody else except you is able to access it, so why would that change after your death - P10

In some cases, such as for P10, quoted above, this is an assumption arrived at naturally, but not described as a consideration that they have made explicitly regarding their post-mortem privacy. One's email account is password protected and typically not accessed by any person but the owner, so there is no initial expectation of a change to this privacy level. This is not necessarily indicative of a belief that the data should in no circumstances be accessed after death - in fact, this participant goes on to express a preference for access to be maintained by a trusted individual. For some, though, treatment of data as private during life is a highly deliberate indicator for continued privacy after death.

> because, for me, anything I've made public is what I want to be public. Anything that's not public, I'm keeping behind closed doors for a good reason - P6

In the case of P6, any data access that is not explicitly encouraged - by way of already being in the public domain, accessible via shared access or delivered as part of a digital legacy plan - is considered to be an invasion of privacy. P6 outlines a range of careful security measures undertaken in protection of this privacy, describing himself as *"extremely paranoid"* in this respect. However, the assumption that this level of privacy is assured and enduring after death may not be entirely accurate. Much of the discussion generated in the consideration of the digital entity cards is illustrative of a lack of confidence in continued privacy after death. Participants describe how changes in security technology may expose their data, or how next of kin or authorities may use legal means to gain access, and therefore *"just assuming that people wouldn't be able to access it afterwards might not be the thing"* (P8). This is accompanied by a lack of trust in large corporations such as Facebook and Google to maintain data privacy or to act in the best interests of the individual. Furthermore, in the event that data access is passed to a trusted data steward, the security of the data may be reduced as a result of the security habits of that individual.

> I randomly generate every password, but then other people I know reuse them, so if their MySpace account gets cracked then that could be immediate access to my inbox - P11

The assumption that data privacy can and should remain the same after death as it was in life is further complicated by various proposed plans and preferences pertaining to the digital entity cards, suggesting that one's death may bring about a change in how some of their data should be handled. For example, in discussing the mobile phone, which presents complications as a potential access point to many different accounts and data items, P2 comments that *"it's weird, you have to be super paranoid with it while you're alive, but obviously once you're dead you want it to be easy to transfer to a really trusted person, but no one else"*. Similarly, data that is restricted during life is not necessarily intended to be restricted indefinitely.

> Facebook, for example – I have a couple of folders that are just from, back from school, when I'd take loads of pictures, and I've made them hidden now, out of no real desire to hide anything, I just don't want them public on my profile at the moment. But if I passed away, then some people might say, 'oh, do you remember [P4] took these pictures, blah blah blah', and they wouldn't be able to access them if I hadn't given somebody the ability to make my account a legacy account - P4

In this way, one's intentions regarding what they do and do not choose to share publicly may be more of a reflection of their identity as they wish it to be perceived at that particular point in time than a statement about the fundamentally private nature of the data itself. As P2 describes it, *"when you're alive, actually you don't want your old photos to be accessible, but when you're dead, probably, it would be nicer if they were"*.

As well as scenarios in which data moves from more private to less private at the point of death, there are also potentially cases where the justification for the existence and consumption of data expires upon an individual's death. In discussing tracked location data, for example that which is routinely gathered by Google, participants voiced their discomfort with the existence of the data at all.

> I'm terrible, because I turned it off deliberately – turned it off because I didn't like the idea of it, but then, when it made using Google Maps slightly harder one day, I turned it back on again, you know - P13

This description of a trade off between a desired level of personal privacy and the value of services received reflects a form of balance between the online service provider and the user. However, upon the death of the user, this balance is disturbed, as the service provider continues to have access to private data but is no longer providing a service in order to do so. The participant thus expresses a desire to have the data deleted after death that *"comes from a slight unease about it existing"* (P13). This example illustrates how one's privacy choices in life may no longer be a true reflection of their continuing privacy preferences after the point of death.

### 4.2 The post-mortem privacy paradox

Over the course of the data collection, much of the planning and discussions revolved around the controlled destruction of private data and the managed sharing of non-sensitive data. The majority of decisions made about the prompt cards were some form of plan, and most participants expressed some clear privacy interests and wishes, both in favour of preservation and of deletion. In discussions about the mobile phone, P2 declared that *"I don't know that there's a clear answer except that you've gotta do something, so you can't hide or ignore it"*. Despite the hypothetical situation in which the participants themselves are no longer living, personal information – particularly that of a sensitive nature – is discussed as something that is to be protected. This may be for the benefit of the individual themselves, for example in the case of P1's preference that *"I don't want people, really, reading through my messages – not that there's anything dodgy or anything"*. It may also be for the benefit of the friends and family left behind, as expressed by P4.

> definitely with like, the family movies and things, you wouldn't ever wanna lose, anyone else to lose access to them - P4

Others expressed a preference for the removal of data whose continued existence is hard to justify. For example, P13 questions whether there is any value in preserving email data, before suggesting that *"maybe it's the OCD in me, wanting to clean things up"*. In consideration of the potential contents of a cloud storage account, which it was agreed might contain a wide variety of both valuable and valueless data, participants found no difficulty in producing examples of data that they considered to be in need of a plan.

> P8: but there might be things like, you know, letters to the bank or the tax people and stuff like that which might be of use
> P10: er, research, I have research data that I would want others to have access to if I'm not able to access it any more because I died, so I would have a plan for that

However, despite the identified need for a comprehensive plan for these items, very few of the participants had engaged with digital legacy plans at any level. Even in the case of a participant with an existing digital will, there were differences between the existing plan and the reasoning exhibited in response to the digital entity card task. P14 argues at first that the email inbox seems to be a "dangerous thing" and should not be included in any plan, before coming to the realisation that this is inconsistent with decisions already made in real life.

> I mean thinking about it, because I initially said hide, in fact I do have a plan in place, so I do have a designated person that will have access to it - P14

In other cases, participants highlighted a certain need for a plan but were forthcoming about having not completed this activity themselves. P8 was clear that there is a *"definite"* need to plan for cryptocurrency, but later shared with the group the absence of a plan for his own cryptocurrency. Discussion during the card-planning activity almost entirely consisted of statements of what would, or should, be done with each of these commonly held digital items, rather than sharing what they have done, or what they will do. In this way, discussion about post-mortem privacy and digital legacy appears to conform to the 'privacy paradox', a known phenomenon wherein the technology users' privacy intentions are not reflected in their actions [24, 36].

In addition to an overall inconsistency between what participants spoke of as in need of planning and their own actions, there was also seen to be a problematic distinction between what they consider to be important in the present and an imagined future in which they have died and hence, no longer have an interest. For example, despite repeatedly expressing a preference for personal information to be *"purged"*, P11 remarks that *"well, I probably won't care, to be honest, because I won't be alive to see whatever someone does with it"*. P4 outlines a preference for illegally obtained files to be destroyed, but follows this up by saying *"there probably won't be any consequences – I'm dead"*. When P9 outlines a concern that an appointed data steward might get hacked and leak his data, other participants suggest that being dead prevents this from impacting him personally.

> P14: you're dead, though, remember
> P8: yeah ... you can't worry about it

This tendency to dismiss privacy as something that is only relevant to the living has the potential to strongly demotivate actions protecting personal privacy after death and direct planning efforts entirely towards assisting those who they will have left behind. However, the argument that nothing matters to the dead can be applied to their loved ones as much as it can be applied to their personal privacy – in both cases, it is what is valuable to a person while they are alive that dictates the nature of any plans made. Nevertheless, despite the various confirmations that participants see value in their post-mortem privacy, discussions of hypothetical scenarios in which they were no longer living were often framed in terms of the interests of their loved ones, with their own values dismissed as no longer relevant.

> I'm sort of in two minds about it, personally, it's either 'who cares, I'm dead', or 'get rid of it', but I guess that suits... any dependents quite well, doesn't it, because you... might not want dependents finding out about stuff that might be private - P11

### 4.3 "It's a bit of a grim thing to be doing"

One of the primary obstacles to encouraging active engagement with digital legacy planning is the nature of the subject matter. In the absence of any knowledge regarding the timing of one's own death, contemplation of what will happen to one's possessions and identity when they die is unlikely to be prioritised.

> nobody knows when they're gonna die really, so it's kinda hard to plan for that eventuality - P4

Much like traditional will-writing, many people are aware that they should have a will, and may intend to create one, but taking the step of doing so requires some kind of motivation or prompt. P13 draws this comparison when asked how people might be encouraged to make plans for their digital legacy by suggesting that *"the same reason anybody would make a will would be the same reasons why people would go through this kind of process"*. This is reflected in existing literature, for example Pfister outlines a common need for a *"motivational trigger"* [40]. However, while there may be the same overall reason for taking this action, the acts of creating a traditional will and creating a digital will are distinct from one another in how that action is completed. In the case of the former, the activity is typically completed in a single session and can be put out of mind afterwards. In the case of the latter, the information contained within the will is much more fluid and in need of more frequent but less exhaustive review.

> [a will] is, in a way it's expensive, yeah, but it's relatively easy, like you go and see someone, they write it, you check it, that's it. Doing all this is quite – I mean, I've done all this, and it takes a long piece of time, and you've got to be continually updating it as well - P14

This continual updating presents a challenge in two ways: it is time consuming, and it also requires constant re-engagement with a subject matter that is not only upsetting, but also somewhat taboo. P13 highlights this in consideration of the idea of collecting digital legacy preferences as a matter of routine, remarking that *"it's a bit of a grim thing to be doing, like 'email', 'username', 'what happens when you die"*. The combination of these two deterrents is enough to prevent most people from engaging, particularly those who, unlike this sample of computer science students and graduates, have little interest in computers or technology.

> [amateur computer users] can't be bothered to faff around with their computer, never mind their accounts, so why would they spend the time doing all of this - P11

In this respect, a wide-reaching solution must have the capacity to make the process much quicker, easier and more accessible to the wider population.

### 4.4 DESIGN CONSIDERATIONS

### 4.5 Sharing vs Bequeathal

A point that was frequently made in both workshops was that, in many cases, the meaningful data that a person might seek to preserve is not of a fundamentally private nature. Photographs, for example, were returned to repeatedly throughout the data collection as an example of data that is highly valuable and whose loss should be avoided. While there may be cases where photographs exist that should not be shared until after death, in the majority of cases there is no detriment to the sharing of these photographs while the owner is still living. Therefore, the task of ensuring continued access to this data after death can in some cases be reframed to become a task that is completed by the owner during life as an act of sharing rather than an act of bequeathal.

> ...if you are using cloud storage you've always got the option to share files or folders, so I feel like it would be better to go down the route of sharing stuff while you're alive - P6

This approach to ensuring the continuation of our valued data shifts the expectation of the individual from needing to explicitly plan for the moment of their death to a more general inclination to share important data. However, while this was recognised as something that *should* be taking place, it was also agreed that *"the reality is a lot of us never do what we should do in terms of organising our data while we're alive"* (P2). The problem, therefore, may lie within the difficulties associated with good data management practices and the obstacle of establishing agreed-upon sharing mechanisms. P11 speaks of his own abandoned attempts to establish such sharing habits with his partner, who *"has an iPhone and refuses to use anything with, like, any other technology"*.

However, despite general agreement that sharing in the present is not utilised to the full extent, in the case of protecting sensitive information, participants described much more active control. In discussing browsing history, participants felt confident in ignoring data, as *"a lot of people are probably already being proactive"* (P8). In this way, appropriate activity during life can prevent there being a need for protective actions within one's digital legacy plan.

> so this is more something that you'd plan already, right now, in the moment, like you... don't wait for yourself to die... to try to delete it - P10

### 4.6 Password managers

Given the central purpose of password managers in these workshops, both as a means of selective recruitment and a focus for discussion, one of the main approaches discussed as a potential solution is closely aligned with the existing functionality of password managers. Password managers exist as a software tool that collects and passively organises credentials for websites and applications, fulfilling a similar task in pulling together a wide variety of unrelated digital activities and assisting in their management. A primary mechanism of this functionality is that as and when new online accounts are created, a decision is made by the user as to whether or not to store that credential in the password manager. Naturally, given the framing of the workshops and the explicit questioning regarding password managers, this functionality of decision-making at the onboarding level tended to be used as the basis for how technology might support a digital legacy process. However, a number of discussion points highlighted challenges associated with the use of password managers to plan for digital legacy.

*4.6.1 The need for granularity.* At the time of writing, password managers sometimes offer some form of emergency contact functionality or credential sharing system. However, access granted via an emergency contact will typically grant access to all credentials, not a subset, meaning that use of such a feature for digital legacy planning usually precludes the ability to control which elements are accessible to the next of kin. Passwords shared during life via a password manager have some degree of granularity, but rely on those they are shared with also being a user of the same software package and fail to accommodate cases where access is only appropriate after the original owner has died. Password managers have an opportunity to provide functionality for a form of digital will, in which – upon proof of death – distinct actions are taken with particular accounts and files at the discretion of the decedent.

*4.6.2 The need for review.* Any given decision regarding a digital entity that is made at the beginning of that entity's existence may change according to the ongoing use of that entity. For example, P1 gives an example of a Reddit account, which through discussion had been agreed might contain sensitive material:

> ... you sign up for a Reddit account and you don't care, and all of a sudden you do weird stuff and then you go, 'ok, actually, I need to hide that', but you set it up as ignore, or whatever - P1

The variance of online accounts in terms of sensitivity in this way has been described in security research, such as Florêncio and Herley's categorisation of accounts according to potential consequences of account compromise, ranging across five levels from "don't care" to "ultra-sensitive" [14]. Incorporating a contextual integrity framing, it is reasonable to expect that the initial categorisation (a new account with little at risk) may change over time due to unforeseen usage of the service, or corresponding lifestyle changes. From a security perspective, this may support review activities regarding elements such as password strength. Equally, from a post-mortem privacy perspective, the decision to share data or credentials may also be best subjected to some form of review. As the number of credentials stored by password managers can grow to be large, this review activity has the potential to become overwhelming. As P4 notes, *"even with a password manager, managing your passwords is a nightmare"*.

*4.6.3 Competition between software packages.* A further complication can arise if more than one software package is used for the same, or similar purposes. Traditional wills are such that only the most recent is valid, however a digital will constructed using a password manager would be under no such limitation. In cases where multiple password managers are in use, there arises a question of which represents the true intent of the deceased, and how to resolve any conflicts. P4 notes that there is already such a conflict between her dedicated password manager and her browser's password manager functionality.

> I don't know what Lastpass holds, and what Google holds, because both of them are incredibly invasive, and it's kinda hard to tell who's in charge, when they're both being like 'I've got the password'- P4

*4.6.4 Additional security measures.* While passwords are an authentication technique ubiquitous across the internet and among hardware devices, there are other mechanisms of determining identity in use that may conflict with a password-only digital legacy plan. Multi-factor authentication (MFA) is a mechanism that is increasingly being used as an additional security measure, requiring two or more separate factors for authentication – typically something a person owns (such as a mobile phone) or that they are (such as a fingerprint). Despite being a recommended security technique, MFA suffers from poor adoption rates [10]. However, our participants, who were more likely to show strong security habits, identified some complications that may come about when attempting to leave a digital legacy using a password manager.

> I think it's difficult, because my phone is essentially my second factor, and I've got a wife, and she would need access to my second factor for some of the important things - P1

In the case of mobile devices acting as the second factor, that device becomes something that should be explicitly planned for as part of this process. This may involve bequeathing the device itself (if it is still in working order) and relevant unlock passwords/pins, or potentially the provision of 2FA recovery codes and a plan to reset or otherwise disable the device. In the case of the former, access is granted to the entire device, which provides unintended access to apps and websites that are signed in, precluding the ability to limit the access granted to the recipient. For the latter, the user is forced to create a complex plan that assumes knowledge on their own part and that of the recipient, and creates difficulties in bequeathing a potentially valuable mobile device, such as a need to provide proof of death in order to reset it. In either case, a consequence of additional security measures is that the ability to simply and effectively leave a digital legacy is impeded.

### 4.7 Email as a primary key

An alternative, or a complement, to password managers as a central gateway point to one's digital life is one's email account(s). If one has kept their registration details for their various accounts, the email is a point of access to a large proportion of the set of online services that a person uses, and is also commonly the primary mechanism for resetting passwords. Consequently, the inclusion of email in a bequeathed list of password credentials may result in unintended access to accounts or data that were not explicitly included, if that email address is the one that is associated with the account.

> if you have someone's phone, and their email address, you can basically become them - P2

As such, access to emails potentially gives a person even more access to the digital possessions of the account holder than a password manager. However, unlike password managers, there is not typically functionality for this information to be curated or controlled in an effective way, leaving the individual with the responsibility of manually removing sensitive aspects of their email contents. The complexity and effectiveness of this task are likely to be strongly impacted by the data management practices of the deceased.

### 4.8 The burden and sustainability of stewardship

Common to the majority of discussed planning solutions was the notion that an individual, or multiple individuals, would be assigned the responsibility of carrying out any given plan. In some cases, this holds a benefit for that person in that the data or credentials passed to them are valuable, whether financially (as in the case of cryptocurrency), logistically (e.g. important documents) or sentimentally (e.g. photographs). In other cases, the role involves the deletion of data or other similar representation on their behalf. This is already problematic when considering that initial burden of responsibility for carrying out planned legacy actions, but becomes even more so when the long-term implications are considered. As mentioned previously, there is an expectation that any private data is kept secure in the long term. There is also a question of sustainability – is it the responsibility of the steward to ensure the preservation of inherited data, even to the point of including it in their own legacy planning?

> first of all, I'm being a burden to somebody else, so there was an initial consent there which needs to be negotiated, and secondly, what happens when they pass on, do they pass it on to somebody else, do they, and sort of eventually, like, your grandchildren have got like fifteen accounts or something they need to manage? - P13

Prior research has highlighted the nature of this burden, addressing the scale of digital collections and the size of the undertaking in attempting to carry out wishes [7, 16]. An ideal digital legacy solution would make provisions for this relationship between an individual and their steward. Rather than simply naming an individual and outlining a plan, some consideration of the effort required on the part of the steward should be encouraged and factored into the plan. Alternatively, the named steward may instead be a part of a professional service, relinquishing loved ones of the bulk of responsibility.

> P10: but you could also do it to a company, so that that company could be, er Gmail for example, but it could already be that it's not a person but a company and they probably do have these
> P8: - or a solicitor, or something like that

Here, the distinction highlighted by Spence between transactional and relational gifting becomes relevant [42]. For relational gifts – to which there is attached some form of meaning or sentimentality – the gift of a digital asset may be less burdensome. The acting of choosing to pass on a particular item can have value for both the giver and the recipient, regardless of the value of the item itself. However, for digital entities to which there is no such value, the role of stewardship may be something that can be assigned or transferred without such delicate considerations. For some legacy plans, automation is described by our participants as the preferred technique. In the case of an online dating profile, there was a preference to make a plan for complete destruction of associated personal data, but only if it could be carried out automatically – in the event that this is not possible, the dating profile should be deliberately excluded from any digital legacy plan. For others, automation was considered as preferable for the sake of protecting loved ones in their time of grief.

> you know it's likely that [the executor is] a close person so they're probably a bit upset that you've passed away. I mean, is it the sort of thing where maybe writing a script that they can just double click an executable that reads a file - P11

In the case of tools such as password managers, automatically executing a given action regarding an account (or generating a template to simplify the process for the steward to do so) is a feasible functionality. For certain actions – in this example, requesting that a dating profile be removed from public view – this may reduce the negative impact on the individuals given the responsibility of taking care of a digital estate. This viewpoint echoes earlier work by [48] in which the needs for personal control and automation are balanced, and the importance of context is advocated.

### 4.9 A legal default

When asked how the wider population might be encouraged to engage in digital legacy planning, the participants in workshop two ultimately arrived at the conclusion that, for most people, this is a difficult or even impossible task. They argued that there is a responsibility for the government to provide a default, opt-out process by which data is legally accessible to one's next of kin.

> what I would see as an ideal that there's a- a default in place, er legal default, that says that all data, all digital data, your direct family gets rights, or one person gets

> direct rights to your digital data the moment you die, unless stated otherwise - P10

Given that, prior to this point in the conversation, many of the preferences stated had been in favour of controlled deletion, this development was somewhat surprising. However, the participants placed value in the simplicity of such a plan from the perspective of an ordinary citizen, making the argument that those people who have a strong preference then have the capacity to opt-out and maintain an appropriate level of privacy. For those who have no such strong opinions, a default such as this supports the needs of their loved ones in accessing and preserving meaningful or valuable digital assets. In the view of these participants, an opt-out approach would be an effective resolution for a wide proportion of the population.

> [the] pensions opt-out means that more people are saving a pension than ever before - P8

Such a solution would introduce a range of privacy, technical, cyber-security and legal questions that are beyond the scope of this paper to discuss (see [19] for an argument against default transmission of email data). However, it also fails to solve some of the main problems associated with digital legacy planning. It would fail to account for many of the preferences outlined throughout the workshop activity, which included not only the preservation or deletion of data, but also the future of various online accounts, treatment of monetary assets and plans for notifying online followers and friends of death. It also assumes some existing knowledge by the next of kin regarding what digital entities exist and why they should consider seeking access to them or to the data contained within them. By introducing a legal default such as this, there is a risk that even less preparation and thought would go into this form of planning, which is illustrated by P13's assertion, gesturing to the assorted digital entity cards, that *"if that legal thing was in place, I'd put all of those in 'ignore' "*.

## 5 DISCUSSION

Our findings challenge the notion that data privacy can, or should, be assumed to remain the same after death. However, even as our participants recognise this and argue for particular actions relating to individual digital items, we identify a reluctance to engage with processes to influence post-mortem privacy outcomes. Much like the broader privacy paradox, we identify a challenge of enabling and encouraging actions to control personal privacy after death – including potentially relinquishing privacy in order to pass on meaningful or valuable digital assets to others. What we have described as the *post-mortem privacy paradox* reflects what we perceive as a gulf between the recognition of post-mortem privacy intentions and the completion of the actions that need to take place in order to support the execution of those intentions. As a result, we suggest that efforts to increase engagement in digital legacy planning may be more compelling when the value to others is stressed, rather than the value for the individual. We argue that for some digital items, the problem of navigating access after death can be avoided by arranging and implementing access for specific individuals during life. In these cases, the challenge may be reframed as one of organisation and communication, rather than one of planning.

Many of the findings we report on highlight individual differences in the values and meaning attached to digital entities. However, without explicit guidance regarding individual items, there is a reliance on an all or nothing approach to post-mortem data access. This produces a tension with existing cybersecurity behaviours – if a person seeks to maintain absolute control over the security of their data during life, they restrict their ability to transfer meaningful or valuable digital assets in the event of their death except by relinquishing their privacy entirely. In doing so, there is not only a cost to their continued privacy, but the recipients, or stewards, of their digital legacy are at risk of receiving an unmanageable amount of data with little to no guidance. In the event of an attempt at continued absolute privacy after death, without the involvement of others there is a risk that the individual's privacy will be diminished over time due to changes in cybersecurity measures and the minimal or absent rights of privacy of the dead.

If digital legacy technologies are to overcome this issue, there is a need for designs that enable granularity of digital legacy planning while supporting continued security of that data. This problem exists in two parts: the provision of usable software solutions, and the social conditions necessary for users to engage with this kind of continuous planning activity. In order for both of these parts to be achieved, there is a need for security tools to take death into consideration. Where tools are designed to pass access from a deceased individual to a living beneficiary or data steward, how can the privacy wishes of the deceased be communicated and implicated? Conversely, where tools are designed for the protection of personal privacy, what considerations are made for the continuation – or adjustment – of that privacy after death?

### 5.1 Limitations & Future Work

Our research is a small-scale exploration, with a focus on a particular sample of more security-aware users. This means that our findings are subject to some degree of bias and should not be taken to be generalisable to the larger population. However, our focus on potential conflicts between usable security and granular digital legacy reveals a design space that is complex and underdeveloped. There is a need to explore in detail the value of the discussed methods and findings with people with a more pressing need to make clear digital legacy plans that do not risk their personal privacy. There is also a need to consider the wider population and identify how best to engage the average user in considering their digital legacy and post-mortem privacy needs. This should extend to examinations of differences across cultures and socio-economic status. From a design perspective, there is a need to consider how best to design password management tools to encourage not only initial consideration of these issues, but also to promote ongoing review and active curation of their digital identities in order to be able to exert control over digital legacy with minimal cost to usability. Finally, there is a need for a more thorough analysis of the extent to which people's post-mortem privacy planning behaviours fail to meet their preferences.

# 6 CONCLUSION

Through analysis of discussion structured around the preferred future of individual digital assets after death, this research has highlighted some of the opportunities and obstacles to the development and uptake of digital legacy processes. Notably, technology users' planning actions are under-developed compared to their values regarding the future of their data and digital assets, and strong security behaviours were identified as a potential impediment to controlling the future of one's data after death. Considering digital entities on an individual basis reveals nuanced variations regarding potential planning activity, however the formation of such plans is hindered by the required data organisation practices, privacy awareness and willingness to engage with the notion of one's own death.

# 7 ACKNOWLEDGMENTS

This work was funded through the EPSRC Centre for Doctoral Training in Digital Civics at Newcastle University (EP/L016176/1). Data, materials and metadata supporting this work are available at https://doi.org/10.25405/data.ncl.13988207.